\documentclass[aps,prb,preprint,superscriptaddress]{revtex4-1}

\usepackage{graphicx}
\usepackage{subfigure}
\usepackage{multirow}

\begin{document}

\title{Correspondence: Reconsidering evidence of shift current in a ferroelectric charge-transfer complex}

\author{Vladimir M Fridkin}
\affiliation{Shubnikov Institute for Crystallography,
	Russian Academy of Sciences, Moscow, Russian Federation}%
\affiliation{Department of Physics,
	Drexel University, Philadelphia, PA 19104, USA}%
\author{Jonathan E. Spanier}
\affiliation{Department of Materials Science \& Engineering,
	Drexel University, Philadelphia, PA 19104, USA}%
\affiliation{Department of Physics,
	Drexel University, Philadelphia, PA 19104, USA}%
\email[ ]{email: spanier@drexel.edu}
\date{\today}

\begin{abstract}
\noindent In Nakamura $\emph{et al.}$~\cite{Nakamura:2017}, the authors present evidence of shift current in the electronic ferroelectric tetrathiafulvalene-$p$-chloranil (TTF-CA).  Since the bulk photovoltaic current in non-centrosymmetric crystals has two contributions, namely the ballistic and shift, we explain why the experimental data and analysis presented by the authors does not permit unambiguous identification of shift current, and is not consistent with the mechanism of shift.
\end{abstract}


\maketitle


\noindent The bulk photovoltaic effect (BPVE) in non-centrosymmetric crystals is an intriguing phenomenon in which photo-generated carriers can be collected in the absence of inhomogeneity in charge or illumination, and photovoltages can greatly exceed the band gap.  In a recent article published in $\emph{Nature Communications}$, Nakamura $\emph{et al.}$~\cite{Nakamura:2017} report that the BPVE can be observed in a molecular ``electronic'' ferroelectric, TTF-CA.  Interestingly the authors report the authors report how their experiments, including observation of an anomalously long distance (200 $\mu$m) of travel of photocarriers constitute a first experimental observation of shift current, pointing to application of such compounds for novel optoelectronic devices based on the experimental observation of shift current.  However we would like to call to the authors' and readers' attention that the interpretation of the nature of this current is in error.

The authors considered the observed photovoltaic current in TTF-CA crystals as shift current.  Moreover they write in the abstract ``Furthermore we reveal that the travel distance of photocarriers exceeds 200 $\mu$m.  These results unveil distinct features of the shift current...''  As shown in our discussion below this conclusion is not valid.

There are two kinds of bulk photovoltaic currents in the crystal lacking a center of inversion.  The first, called ballistic, is connected with diagonal elements of the density matrix and is caused by the band transport of carriers with asymmetric momentum distribution~\cite{Sturman:1992}.  The ballistic current reveals the Hall effect and a large mobility of non-thermalized carriers.  Practically all ferroelectric and piezoelectric crystals reveal ballistic current, which exists both in intrinsic and extrinsic optical regions.  There exists another contribution to the blk photovoltaic current, one that is connected with off-diagonal density matrix elements~\cite{Sturman:1992, Belincher:1980,Belincher:1982,Belincher:1988,Young:2012a,Young:2012b}.  This contribution was termed shift current because it is due to the shift of carriers in real space when they under go any quantum transition.  By contrast to the ballistic, the shift photovoltaic current is not connected with carrier band transport and it is insensitive to external magnetic field.  The relaxation time of ballistic carriers is determined by the non-thermalized carriers lifetime ($10^{-12}-10^{-13}$ sec.).  The theory shows that shift current $j^{sh}$ is in principle the same order of magnitude as ballistic $j^{bal}$, but could be much smaller, depending on the product of wavevector $k$ and lattice constant $a$~\cite{Sturman:1992,Belincher:1980,Belincher:1982,Belincher:1988}. Until now shift current has not been observed experimentally, and separation of $j^{bal}$ and $j^{sh}$ has not been achieved experimentally.  Nevertheless, in Nakamura $\emph{et al.}$~\cite{Nakamura:2017} ballistic current is apparently ignored without justification.

The shift collection length presented in the paper $R$ $\approx$ 200 $\mu$m contradicts the well known theoretical data~\cite{Sturman:1992} by three to four orders of magnitude.  The shift $R$ determines the value of shift current $j^{sh}$:~\cite{Sturman:1992} [N.B. p. 83. Eq. (2.94) therein]

\begin{equation}
j^{sh} = e\frac{\alpha I}{\hbar\omega}R
\end{equation}

\noindent where $j^{sh}$ is the density of the shift current, $e$ the elementary charge, $\alpha$ the absorption coefficient, $I$ the light intensity, and $\hbar\omega$ the photon energy. From the authors' data $R = 2\times10^{-2}$ cm, $I$ = 0.1 W/cm$^{-2}$, $\hbar\omega \approx 3$ eV. 
Optical absorption is not indicated, but from the thickness specified (0.2 mm) and assuming homogeneous absorption, we estimate $\alpha$ $\lesssim$ 50 cm$^{-1}$.  From Eq. (1) we have $j^{sh} \approx$ 30 mA cm$^{-2}$.  This is four orders of magnitude larger than the current density obtained by the authors (1.6 $\mu$A cm$^{-2}$).  Therefore attributing the 0.2 mm ``anomalously long travel distance of photocarriers'' collection length to shift is simply not realistic.

One might imagine a situation, however improbable, in which the collected current, in principle, arises from many successive shift mechanism events, $\emph{i.e.}$, successive processes of photo-excitation and entangled electron-photon phase coherence, relaxation and de-phasing, and subsequent re-excitation, resulting in collection on the scale of 0.2 mm.  This is three to four orders of magnitude larger than the theoretically predicted shift vector magnitude.  

Shift current, by definition, only involves shift of carriers under illumination.  Accordingly the possibility that current collected from a single local (10 $\mu$m-wide) illuminated spot at some distance (200 $\mu$m) from the electrical contact (as shown in Fig. 4a) somehow permits photo-generated carriers to travel thousands of shift distances in region(s) not under illumination is nonphysical, fundamentally incompatible with the definition of shift, which has no meaning in the absence of illumination.  The authors have written ``...we reveal that the travel distance of photocarriers exceeds 200 $\mu$m.  These results unveil distinct features of the shift current...''  Absent a proper explanation, the authors' attribution to shift from scanning the local illumination spot along the $c$-axis reflects a significant, fundamental misunderstanding of the theory and physical nature of shift.

Further, the authors present an ultrafast time response characteristic (Fig. 5) as evidence of shift, which is also misleading.  They have written ``It is known that the shift current shows an ultra-fast response for pulse light, whose temporal waveform follows that of a pump pulse. Figure 5a presents the transient photocurrent responses for pulse light in the I phase measured using a 130-fs-width pulse laser as a light source. They initially exhibit a pulse current with fast decay and subsequent oscillating components. The first component of the photocurrent can be attributed to the shift current.''  They also remark ``In addition to applications for energy harvesting, the ultra-fast response time of the shift current is considered useful for many devices including sensors and actuators.''  In fact the ballistic timescale (timescale of thermalization) is on the same order, $\emph{i.e.}$, 10$^{-12}$-10$^{-13}$ sec. as this ultrafast response.  Thus the data in Fig 5 cannot be taken as evidence of shift.\\
\indent We would like to call to the readers' attention the following additional points.  Perhaps the authors are unaware that both the ballistic and shift phenomena are explained by the same tensor of the third rank. To explain the observed anisotropy of the photovoltaic current, described by the authors as a ``striking feature'', one needs simply to know the symmetry of the third rank tensor which is responsible for this feature.  However the point group of the crystal was not provided.  Also, we would like to point out that the authors explain their experimental observation within the context of the Rice-Mele one-dimensional model~\cite{Rice:1982}.  However, the authors' attribution of a phase transition at $T_c$ = 80 K in their so-called 1D ferroelectric chain is in contradiction with the Landau theory~\cite{Landau} which states that a phase transition cannot exist in a one-dimensional system.\\
\indent In conclusion, the ballistic and shift currents in non-centrosymmetric crystals now attract great interest in connection with photovoltaics~\cite{Young:2012a,Spanier:2016,Tan:2016,Rappe:2017}.  The assignment of the observed long distance of photo-generated carrier transport ($\approx$ 200 $\mu$m) to shift current is, in our view, not only unrealistic and misleading: the current differs by orders of magnitude from that which is predicted for shift, and there has been apparently no consideration given to the fact that there are two contributions to the bulk photovoltaic effect.  We encourage the authors to carefully reconsider their claim that their experimental observations in a ferroelectric charge-transfer complex are explained by shift current.

\end{document}